\documentclass[
12pt,
times
]{elsarticle}

\usepackage[colorlinks=true, breaklinks=true 
]{hyperref}

\usepackage{graphicx} 

\usepackage{color}
\usepackage{lineno}

\journal{European Journal of Physics}

\begin{document}

\begin{frontmatter}

\title{Expanding the simple pendulum's rotation solution in action-angle variables}

\author[lara]{ Martin Lara\fnref{yo} \corref{cor1}  }
\ead{mlara0@gmail.com}

\author[ferrer]{ Sebasti\'an Ferrer\fnref{el} } 
\ead{sferrer@um.es}

\fntext[yo]{ Instituto de Ciencia y Technologia, Universidade Federal de S\~ao Paulo }
\fntext[el]{ Departamento de Matem\'aticas y Computaci\'on, Universidad de Murcia }

\address[lara]{Rua Talim, 330, 12231-280 S\~ao Jos\'e dos Campos, SP, Brazil}
\address[ferrer]{Campus de Espinardo, 30071 Espinardo, Murcia, Spain}

\cortext[cor1]{Corresponding author}

\date{\today}

\begin{abstract}
Integration of Hamiltonian systems by reduction to action-angle variables has proven to be a successful approach.
However, when the solution depends on elliptic functions the transformation to action-angle variables may need to remain in implicit form.
This is exactly the case of the simple pendulum, where in order to make explicit the transformation to action-angle variables one needs to
resort to nontrivial expansions of special functions and series reversion.
Alternatively, it is shown that the explicit expansion of the transformation to action-angle variables can be constructed directly, and that this direct construction leads naturally to the Lie transforms method, in this way avoiding the intricacies related to the traditional expansion of elliptic functions.
\end{abstract}


\end{frontmatter}

\section{Introduction} 

The simple gravity pendulum, a massless rod with a fixed end and a mass attached to the free end, is one of the simplest integrable models. However, it provides a useful didactic system that may be available in most students laboratories to be used in different levels of physics \cite{NelsonOlssonAJP1986}. Besides, this physical model serves as the basis for investigating many different phenomena which exhibit a variety of motions including chaos \cite{BakerBlackburn2005}, and has the basic form that arises in resonance problems \cite{FerrazMello2007}. The case of small oscillations about the stable equilibrium position is customarily studied with linearized dynamics, and was for a long time the basis for implementing traditional timekeeping devices.
\par

The dynamical system is only of one degree of freedom, but the motion may evolve in different regimes, and one must resort to the use of special functions to express its general solution in closed form \cite{Ochs2011}. In fact, this dynamical model is commonly used to introduce the Jacobi elliptic functions \cite{Brizard2009}.
\par

The traditional integration provides the period of the motion as a function of the pendulum's length and the initial angle \cite{LimaAJP2010}. Alternatively, the solution can be computed by Hamiltonian reduction, a case in which action-angle variables play a relevant role  \cite{Brizard2013}. In particular, they are customarily accepted as the correct variables for finding approximate solutions of almost integrable problems by perturbation methods \cite{Arnold1989}.
\par

When the closed-form solution of an integrable problem is expressed in terms of standard functions, the transformation to action-angle variables can be made explicitly in closed form, as, for instance, in the case of the harmonic oscillator. However, if the solution relies on special functions, whose evaluation will depend on one or more parameters in addition to the function's argument, the action-angle variables approach may provide the closed-form solution in an implicit form. This fact does not cause trouble when evaluating the solution of the integrable problem, but may deprive this solution of some physical insight. Besides, in usual perturbation methods the disturbing function must be expressed in the action-angle variables of the integrable problem, thus making necessary to expand the (implicit) transformation to action-angle variables as a Fourier series in the argument of the special functions. These kinds of expansions are not trivial at all, and finding them may be regarded as a notable achievement \cite{Sadov1970,SadovRuso1970,Kinoshita1972}.
\par

In the case of elliptic functions, the normal way of proceeding is to replace them by their definitions in terms of Jacobi theta functions, which in turn are replaced by their usual Fourier series expansion in trigonometric functions of the elliptic argument, with coefficients that are powers of the elliptic nome \cite{ByrdFriedman1971,Lawden1989}. This laborious procedure is greatly complicated when the modulus of the elliptic function remains as an implicit function of the action-angle variables, a case that requires an additional expansion and the series reversion of the resulting power series.
\par

On the other hand, the expansion of the transformation to action-angle variables can be constructed directly. 
Indeed, the assumption that the transformation equations of the solution are given by Taylor series expansions, with the requirement that this transformation reduces the pendulum Hamiltonian to a function of only the momenta, and the constrain that the transformation be canonical leads naturally to the Lie transforms procedure \cite{Hori1966,Deprit1969}. The procedure is illustrated here for the rotation regime of the simple pendulum \cite{PercivalRichards1982,SussmanWisdom2001}. Indeed, after rearranging the Lie transforms solutions as a Fourier series, we check that both kinds of expansions match term for term. Therefore, there is no need of making any numerical experiment to show the convergence of the Lie transforms solution for the pendulum's rotation regime \citep{SussmanWisdom2001}.
\par

In the case of the pendulum's oscillation regime the series expansion of the closed form solution as a Fourier series in the action-angle variables still can be done. However, these expansions provide a relation between an angle whose oscillations are constrained to a maximum elongation, and an angle that rotates, for this reason not fulfilling the condition required by the Lie transforms method that original and transformed variables are the same when the small parameter vanishes. On the contrary, the Lie transforms solution for the oscillation regime is commonly approached after reformulating the pendulum Hamiltonian as a perturbed harmonic oscillator by making use of a preliminary change of variables to Poincar\'e canonical variables. This case is well documented in the literature \cite{LichtenbergLieberman1992,PercivalRichards1982} and is no tackled here. 

\section{Hamiltonian reduction of the simple pendulum}

The Lagrangian of a simple pendulum of mass $m$ and length $l$ under the only action of the gravity acceleration $g$ is written 
$\mathcal{L}=T-V$ where, noting $\theta$ the angle with respect to the vertical direction, $V=mgl(1-\cos\theta)$ is the potential energy 
and $T=(1/2)I\omega^2$ is the kinetic energy, where $I=ml^2$ is the pendulum's moment of inertia and $\omega=\dot\theta$, where the over 
dot means derivation with respect to time.
\par

The conjugate momentum to $\theta$ is given by
\[
\Theta=\frac{\partial\mathcal{L}}{\partial\dot\theta}=ml^2\dot\theta
\]
That is, $\Theta=I\omega$ is the angular momentum.

Hence, the usual construction of the Hamiltonian $\mathcal{H}=\Theta\,\dot\theta(\Theta)-\mathcal{L}(\theta,\dot\theta(\Theta))$ gives
\begin{equation} \label{pendulum:ham}
\mathcal{H}=\frac{1}{2}\frac{\Theta^2}{ml^2}+mgl(1-\cos\theta),
\end{equation}
which represents the total energy for given initial conditions $\mathcal{H}(\theta_0,\Theta_0)=E$. Depending on the energy value the 
pendulum may evolve in three different regimes:
\begin{itemize}
\item $0\le{E}<2mgl$, the oscillation regime, with a fixed point of the elliptic type at $E=0$ ($\Theta=0$, $\theta=0$).
\item $E=2mgl$ $\Rightarrow$ $\Theta=\pm2ml^2\sqrt{g/l}\cos(\theta/2)$, the separatrix, with fixed points of the hyperbolic type 
$\Theta=0$, $\theta=\pm\pi$.
\item $E>2mgl$, the rotation regime
\end{itemize}
\par

\subsection{Phase space}

For each energy manifold $\mathcal{H}=E$, the Hamiltonian (\ref{pendulum:ham}) has a geometric interpretation as a parabolic cylinder 
$\mathcal{P}_E=\{ (x,y,z) \,|\, z^2-2y=h\}$, with
\[
x=l\sin\theta, \qquad y=l\cos\theta, \qquad z=\frac{\Theta}{mlg^{1/2}},
\]
and
\[
h=\frac{2E}{mg}-2l=\mathrm{const.}
\]
\par

\begin{figure}[htbp]
\centering
\includegraphics[scale=0.75]{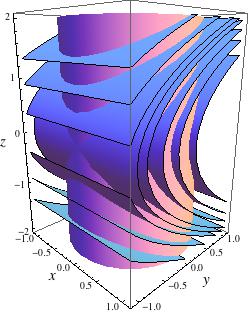}\quad\includegraphics[scale=0.75]{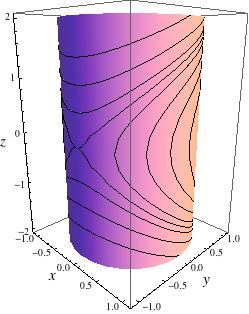}
\caption{Geometric construction of the pendulum's phase space. Parabolic cylinders $\mathcal{P}_E=\{ (x,y,z) \,|\, z^2-2y=\mathrm{const.}\}$.}
\label{f:pendulum:cylinder}
\end{figure}

Besides, the constraint $x^2+y^2=l^2$ makes that the phase space of the simple pendulum is realized by the intersection of parabolic 
cylinders, given by the different energy levels of the Hamiltonian (\ref{pendulum:ham}), with the surface of the  cylinder 
$\mathcal{C}=\{ (x,y,z) \,|\, x^2+y^2=l^2\}$ of radius $l$. This geometric interpretation of the phase space is illustrated in 
Fig.~\ref{f:pendulum:cylinder}.
\par

\begin{figure}[htbp]
\centering
\includegraphics[scale=1.2]{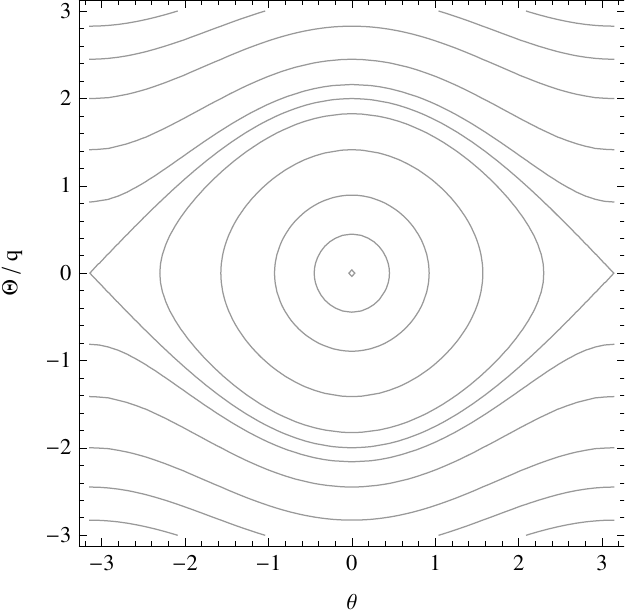}
\caption{Phase space of the simple pendulum ($q=\sqrt{2m^2gl^3}$).}
\label{f:pendulum:phsp}
\end{figure}

Alternatively, typical trajectories on the cylinder are displayed by means of simple contour plots of the Hamiltonian (\ref{pendulum:ham}), 
as illustrated in Fig.~\ref{f:pendulum:phsp} where the trajectories are traveled from left to right for positive heights ($\Theta>0$) and 
from right to left for negative heights ($\Theta<0$). 
\par

The traditional solution to the motion of the pendulum is approached by the direct integration of Hamilton equations
\[
\dot\theta=\frac{\partial\mathcal{H}}{\partial\Theta}=\frac{\Theta}{ml^2},
\qquad
\dot\Theta=-\frac{\partial\mathcal{H}}{\partial\theta}=-mlg\sin\theta,
\]
which are usually written as a single, second order differential equation
\begin{equation} \label{pendulum:flow}
\ddot\theta+\frac{g}{l}\sin\theta=0.
\end{equation}
\par

Note that the summand $mgl$ in Eq.~(\ref{pendulum:ham}) is a constant term that does not affect the dynamics derived from Hamilton equations. 
Hence, this term is sometimes neglected, which implies a trivial displacement of the energy by a constant level. Besides, the Hamiltonian 
can be scaled by the pendulum's moment of inertia to show that it only depends on a relevant parameter. Nevertheless, the dimensional 
parameters are maintained in following derivations, because, in addition to the physical insight that they provide, the simple and immediate 
test of checking dimensions is very useful in verifying the correctness of the mathematical developments.
\par

\subsection{Hamiltonian reduction}

Alternatively to the classical integration of Eq.~(\ref{pendulum:flow}), the flow can be integrated by Hamiltonian reduction, finding a 
transformation of variables
\begin{equation} \label{pendulum:transformation}
\mathcal{T}:(\theta',\Theta')\rightarrow(\theta,\Theta),
\end{equation}
such that the Hamiltonian (\ref{pendulum:ham}) when expressed in the new variables is only a function of the new momentum. Namely,
\begin{equation} \label{pendulum:reduction}
\mathcal{T}:\mathcal{H}\equiv\mathcal{H}(\theta(\theta',\Theta'),\Theta(\theta',\Theta'))=\Phi(-,\Theta').
\end{equation}
The solution of Hamilton equations in the new variables
\[
\dot\theta'=\frac{\partial\Phi}{\partial\Theta'}, 
\qquad
\dot\Theta'=-\frac{\partial\Phi}{\partial\theta'}=0,
\]
is trivial
\begin{equation} \label{pendulum:sol}
\Theta'=\mathrm{const},
\qquad
\theta'=\theta'_0+\omega't,
\end{equation}
where the frequency $\omega'=\partial\Phi/\partial\Theta'$ only depends on $\Theta'$ and, therefore, is constant. 
Plugging Eq.~(\ref{pendulum:sol}) into the transformation $\mathcal{T}$ in Eq.~(\ref{pendulum:transformation}) will give the time solution 
in the original phase space $(\theta,\Theta)$.
\par

The Hamiltonian reduction in Eq.~(\ref{pendulum:reduction}) can be achieved by the Hamilton-Jacobi method \cite{GoldsteinPooleSafko2001},
in which the canonical transformation $\mathcal{T}$ is derived from a generating function $\mathcal{S}=\mathcal{S}(\theta,\Theta')$ in 
mixed variables, the ``old'' coordinate and the ``new'' momentum, such that the transformation is given by
\begin{equation} \label{pendulum:canonicalpQ}
\theta'=\frac{\partial\mathcal{S}}{\partial\Theta'},\qquad 
\Theta=\frac{\partial\mathcal{S}}{\partial\theta}.
\end{equation}
\par

The functional expression of $\Theta$, given by the last of Eq.~(\ref{pendulum:canonicalpQ}) is replaced in the 
Hamiltonian (\ref{pendulum:ham}) to give the partial differential equation
\begin{equation} \label{pendulum:PDE}
\frac{1}{2ml^2}\left(\frac{\partial\mathcal{S}}{\partial\theta}\right)^{\!2}+2mgl\sin^2(\theta/2)=\Phi(\Theta'),
\end{equation}
where the trigonometric identity $1-\cos\theta=2\sin^2(\theta/2)$ has been used.
\par

The Hamilton-Jacobi equation (\ref{pendulum:PDE}) is a partial differential equation that can be solved by quadrature
\begin{equation} \label{pendulum:S}
\mathcal{S}=\sqrt{2ml^2}\int\sqrt{\Phi(\Theta')-2mgl\sin^2(\theta/2)}\,\mathrm{d}\theta.
\end{equation}
But, in fact, the generating function $\mathcal{S}$ does not need to be computed because the transformation (\ref{pendulum:canonicalpQ}) 
only requires the partial derivatives of Eq.~(\ref{pendulum:S}). Thus,
\begin{eqnarray}  
\label{pendulum:thetap}
\theta'&=& 
\sqrt{2ml^2}\,\frac{\partial\Phi}{\partial\Theta'}\int_{0}^\theta\frac{\mathrm{d}\vartheta}{2\sqrt{\Phi(\Theta')-2mgl\sin^2(\vartheta/2)}},
\\
\label{pendulum:Theta}
\Theta &=& \sqrt{2ml^2}\sqrt{\Phi(\Theta')-2mgl\sin^2(\theta/2)},
\end{eqnarray}
where $\vartheta=0$ corresponds to the upper extreme of the square root in Eq.~(\ref{pendulum:thetap}).
\par

Remarkably, the quadrature in Eq.~(\ref{pendulum:thetap}) can be solved without need of choosing the form of the new Hamiltonian in 
advance \cite{SussmanWisdom2001,FerrerLara2010b}.
\par

\subsection{Rotation regime: $\Phi(\Theta')>2mgl$}

Equation (\ref{pendulum:thetap}) is reorganized as
\[
\theta'=\sqrt{\frac{l}{g}}\,k\,\frac{\mathrm{d}\Phi}{\mathrm{d}\Theta'}\,\mathrm{F}(\phi,k^2),
\]
where the partial differentiation has been replaced by the total derivative in view of the single dependency of $\Phi$ on $\Theta'$, and
\[
\mathrm{F}(\phi,k^2)=\int_0^\phi\frac{\mathrm{d}\theta}{\sqrt{1-k^2\sin^2\theta}},
\]
is the incomplete elliptic integral of the first kind of \emph{amplitude}
\begin{equation} \label{pendulum:argument}
\phi=\theta/2,
\end{equation}
and \emph{elliptic modulus}
\begin{equation} \label{pendulum:krotation}
k=\sqrt{\frac{2mgl}{\Phi(\Theta')}}<1.
\end{equation}
\par

Therefore, the rotational motion of the simple pendulum is solved by the transformation in mixed variables
\begin{eqnarray} \label{pendulum:mixed}
\theta' &=& \sqrt{\frac{l}{g}}\,k\,\frac{\mathrm{d}\Phi}{\mathrm{d}\Theta'}\,\mathrm{F}(\phi,k^2),
\\ \label{pendulum:Qc}
\Theta &=& 2m\,l^2\sqrt{\frac{g}{l}}\,\frac{1}{k}\,\sqrt{1-k^2\sin^2\phi},
\end{eqnarray}
which is, in fact, a whole \emph{family} of transformations parameterized by $\Phi$. For instance, a ``simplifying'' option could be 
choosing $\Phi\equiv\Theta'^2/(8ml^2)$, which results in
\begin{eqnarray} \label{pendulum:theta0}
\sin\frac{\theta}{2} &=& \mathrm{sn}(\theta',k^2),
\\ \label{pendulum:Theta0}
\frac{\Theta}{2\sqrt{m^2gl^3}} &=& \frac{1}{k}\,\mathrm{dn}(\theta',k^2),
\end{eqnarray}
where $k=4\sqrt{m^2gl^3}/\Theta'$, from Eq.~(\ref{pendulum:krotation}). Note that $\phi=\mathrm{am}(\theta',k^2)$, where the Jacobi 
amplitude $\mathrm{am}$ is the inverse function of the elliptic integral of the first kind,  and $\mathrm{sn}$ and $\mathrm{dn}$ are the 
Jacobi sine amplitude and delta amplitude functions, respectively. 
\par


From the definition of $k$ in Eq.~(\ref{pendulum:krotation}), one may express the reduced Hamiltonian in the \emph{standard} form
\begin{equation} \label{pendulum:StandardHam}
\Phi=2mgl\frac{1}{k^2}.
\end{equation}
Then, by simple differentiation,
\begin{equation}  \label{pendulum:dPhidThetap}
\frac{\mathrm{d}\Phi}{\mathrm{d}\Theta'}=-4mgl\frac{1}{k^3}\,\frac{\mathrm{d}k}{\mathrm{d}\Theta'}.
\end{equation}
which turns Eq.~(\ref{pendulum:mixed}) into
\begin{equation}
\theta'= -4mgl\sqrt{\frac{l}{g}}\,\frac{1}{k^2}\,\frac{\mathrm{d}k}{\mathrm{d}\Theta'}\,\mathrm{F}(\phi,k^2).
\end{equation}

\section{Action-angle variables}

Note that, in general, the new variables $(\theta',\Theta')$ defined by the mixed transformation in Eqs.~(\ref{pendulum:mixed}) 
and (\ref{pendulum:Qc}) may be of a different nature than the original variables $(\theta,\Theta)$: an angle and angular momentum, 
respectively. Indeed, the new Hamiltonian choice leading to Eqs.~(\ref{pendulum:theta0})--(\ref{pendulum:Theta0}) assigns $\theta'$ the 
dimension of time. However, choosing a transformation that preserves the dimensions of the new variables $(\theta',\Theta')$ as angle and 
angular momentum ---the so-called transformation to action-angle variables--- may provide a deeper geometrical insight of the solution and, 
furthermore, is specially useful in perturbation theory. Alternative derivations to the one given below can be found in the literature \cite{FerrazMello2007}.
\par

Hence, the requirement that $\theta'=\theta'(\theta,\Theta)$ be an angle, given by the condition 
$\oint\mathrm{d}\theta'=2\pi$ \cite{Arnold1989}, is imposed to Eq.~(\ref{pendulum:mixed}). That is, when $\theta$ completes a period along 
an energy manifold $\mathcal{H}(\theta,\Theta)=E$ then $\theta'$ must vary between $\theta'_0$ and $\theta'_0+2\pi$.
\par

Note in Eq.~(\ref{pendulum:argument}) that when $\theta$ evolves between $0$ and $2\pi$, $\phi=\phi(\theta)$ evolves between $0$ and $\pi$. 
Therefore, using Eq.~(\ref{pendulum:mixed}), the angle condition reads
\[
2\pi= \sqrt{\frac{l}{g}}\,k\,\frac{\mathrm{d}\Phi}{\mathrm{d}\Theta'}\left[\mathrm{F}(\pi,k^2)-\mathrm{F}(0,k^2)\right],
\]
where
\[
\mathrm{F}(0,k^2)=0,\qquad \mathrm{F}(\pi,k^2)=2\mathrm{F}(\pi/2,k^2)=2\mathrm{K}(k^2).
\]
Hence,
\begin{equation}\label{pendulum:PhidP}
\frac{\mathrm{d}\Phi}{\mathrm{d}\Theta'}=\sqrt{\frac{g}{l}}\,\frac{\pi}{k\,\mathrm{K}(k^2)},
\end{equation}
which is replaced in Eq.~(\ref{pendulum:mixed}) to give
\begin{equation} \label{ppendulum:thetap}
\theta'=\frac{\pi}{\mathrm{K}(k^2)}\,\mathrm{F}(\phi,k^2).
\end{equation}
\par

On the other hand, by eliminating $\mathrm{d}\Phi/\mathrm{d}\Theta'$ between Eqs.~(\ref{pendulum:dPhidThetap}) and (\ref{pendulum:PhidP}) 
results in separation of variables. Then, $\Theta'$ is solved by quadrature to give
\begin{equation} \label{ppendulum:pc}
\Theta'=\frac{4}{\pi}\,m\,l^2\,\sqrt{\frac{g}{l}}\,\frac{1}{k}\,\mathrm{E}(k^2),
\end{equation}
where $\mathrm{E}(k^2)=\mathrm{E}(\pi/2,k^2)$ is the \emph{complete} elliptic integral of the second kind, and
\begin{equation} \label{ppendulum:ellipticE}
\mathrm{E}(\phi,k^2)=\int_0^\phi\sqrt{1-k^2\sin^2\theta}\,\mathrm{d}\theta,
\end{equation}
is the \emph{incomplete} elliptic integral of the second kind.
\par

It deserves noting that the elliptic modulus cannot be solved explicitly from Eq.~(\ref{ppendulum:pc}), and hence the reduced 
Hamiltonian $\Phi$ must remain as an implicit function of $\Theta'$ in the standard form of Eq.~(\ref{pendulum:StandardHam}). However, the 
rotation frequency of $\theta'$ is trivially derived from Hamilton equations by using Eq.~(\ref{pendulum:PhidP}), namely
\begin{equation} \label{pendulum:nr}
n_{\theta'}=\frac{\mathrm{d}\theta'}{\mathrm{d}t}=\frac{\mathrm{d}\Phi}{\mathrm{d}\Theta'}=\sqrt{\frac{g}{l}}\,\frac{\pi}{k\,\mathrm{K}(k^2)}.
\end{equation}
\par

In summary, starting from $(\theta,\Theta)$, corresponding action-angle variables $(\theta',\Theta')$ are computed from the algorithm:
\begin{enumerate}
\item compute $E=\mathcal{H}(\theta,\Theta)$ from Eq.~(\ref{pendulum:ham});
\item compute $k$ from Eq.~(\ref{pendulum:krotation}) with $\Phi=E$;
\item compute $\Theta'$ from Eq.~(\ref{ppendulum:pc}), and $\theta'$ from Eq.~(\ref{ppendulum:thetap}), where $\phi=\theta/2$.
\end{enumerate}
\par

On the other hand, starting from $(\theta',\Theta')$, the inverse transformation is computed as follows.
\begin{enumerate}
\item solve the implicit equation (\ref{ppendulum:pc}) for $k$;
\item invert the elliptic integral of the first kind in Eq.~(\ref{ppendulum:thetap}) to compute
\[
\phi=\mathrm{am}\!\left(\pi^{-1}\mathrm{K}(k^2)\,\theta',k^2\right);
\]
\item $\theta=2\phi$, and $\Theta$ is evaluated from Eq.~(\ref{pendulum:Qc}).
\end{enumerate}

\section{Series expansions}

In spite of the evaluation of elliptic functions is standard these days, using approximate expressions based on trigonometric functions 
will provide a higher insight into the nature of the solution and may be accurate enough for different applications. 
\par

Thus, Eqs.~(\ref{ppendulum:thetap}) and (\ref{pendulum:Qc}) are written
\begin{eqnarray} \label{pendulum:qk}
\theta &=& 2\,\mathrm{am}(u,k^2),
\\ \label{pendulum:Qk}
\Theta &=& 2ml^2\sqrt{\frac{g}{l}}\,\frac{1}{k}\,\mathrm{dn}(u,k^2),
\end{eqnarray}
which is formally the same transformation as the one in Eqs.~(\ref{pendulum:theta0}) and (\ref{pendulum:Theta0}), except for the argument
\[
u\equiv\mathrm{F}(\phi,k^2)=\frac{\mathrm{K}(k^2)}{\pi}\theta',
\]
which now fulfils the requirement that $\theta'$ be an angle.
\par

The elliptic functions in Eqs.~(\ref{pendulum:qk}) and (\ref{pendulum:Qk}) are expanded using standard relations,\footnote{See \texttt{http://dlmf.nist.gov/22.11E3} and \texttt{http://dlmf.nist.gov/22.16.E9} for Eqs.~(\ref{pendulum:dns}) and (\ref{pendulum:ams}), respectively.}
\begin{eqnarray} \label{pendulum:dns}
\mathrm{dn}(u,k^2) &=&
\frac{\pi}{\mathrm{K}(k^2)}\left[\frac{1}{2}+2\sum_{n=1}^{\infty}\frac{q^{n}}{1+q^{2n}}\cos(n\theta')\right]\!,\quad
\\ \label{pendulum:ams}
\mathrm{am}(u,k^2) &=&
\frac{1}{2}\theta'+2\sum_{n=1}^{\infty}\frac{q^{n}}{n(1+q^{2n})}\sin(n\theta'),
\end{eqnarray}
where the \emph{elliptic nome} $q$ is defined as
\begin{equation} \label{pendulum:nome}
q= \exp\left[-\pi\,\mathrm{K}(1-k^2)/\mathrm{K}(k^2)\right],
\end{equation}
and $k$ must be written explicitly in terms of $\Theta'$. This is done by expanding $\Theta'$ as given by Eq.~(\ref{ppendulum:pc}) in power 
series of $k$, viz.
\begin{eqnarray*}
\Theta' &=& \frac{2\sqrt{g l^3 m^2}}{k}\left[
1-\frac{1}{4}k^2-\frac{3}{64}k^4-\frac{5}{256}k^6-\frac{175}{16384}k^8 \right. \\ && \left. 
-\frac{441}{65536}k^{10}-\frac{4851}{1048576}k^{12}+O(k^{13})\right].
\end{eqnarray*}
Then, by series reversion,
\begin{equation} \label{pendulum:kThetap}
k=2\sqrt{\epsilon}\left[1-\epsilon
+\frac{5}{4}\epsilon^2
-\frac{7}{4}\epsilon^3
+\frac{161}{64}\epsilon^4
-\frac{239}{64}\epsilon^5 +O(\epsilon^6) 
\right],
\end{equation}
where the notation
\begin{equation} \label{pendulum:epsilonThetap}
\epsilon=\frac{m^2gl^3}{\Theta'^2},
\end{equation}
has been introduced. Needles to say that $\epsilon$ must be small in order to the expansions make sense.
\par

Once the required operations have been carried out, the transformation to action-angle variables given by Eqs.~(\ref{pendulum:qk}) 
and (\ref{pendulum:Qk}) is written as the expansion
\begin{eqnarray} \nonumber
\theta &=& \theta'
+\left(1+\frac{11}{16}\epsilon^2+\frac{247}{256}\epsilon^4\right)\epsilon\sin\theta' \\ \nonumber
&& +\left(\frac{1}{8}+\frac{3}{16}\epsilon^2\right)\epsilon^2\sin2\theta'
+\left(\frac{1}{48}+\frac{3}{64}\epsilon^2\right)\epsilon^3\sin3\theta' \\ \label{pendulum:qseries}
&& +\frac{1}{256} \epsilon^4 \sin4\theta'
+\frac{1}{1280}\epsilon^5 \sin5\theta' +O(\epsilon^6),
\\ [1ex] \nonumber
\Theta &=& \Theta'\left[
1-\frac{1}{2}\epsilon^2-\frac{15}{32}\epsilon^4
+\left(1+\frac{3}{16}\epsilon^2+\frac{39}{256}\epsilon^4\right)\epsilon\cos\theta' \right. \\ \nonumber
&&
+\left(\frac{1}{4}+\frac{1}{4}\epsilon^2\right)\epsilon^2\cos2\theta'
+\left(\frac{1}{16}+\frac{7}{64}\epsilon^2\right)\epsilon^3\cos3\theta' \\ \label{pendulum:Qseries}
&& \left.
+\frac{1}{64}\epsilon^4\cos4\theta'
+\frac{1}{256}\epsilon^5\cos5\theta'+O(\epsilon^6)
\right].
\end{eqnarray}
\par

Proceeding analogously, the standard Hamiltonian in Eq.~(\ref{pendulum:StandardHam}) is expanded like
\begin{equation} \label{pendulum:hamSeries}
\Phi=\frac{\Theta'^2}{2l^2m}\left(1+\frac{1}{2}\epsilon^2+\frac{5}{32}\epsilon^4+\frac{9}{64}\epsilon^6+\dots\right).
\end{equation}
\par

\section{Action-angle variables expansions by Lie transforms}

In spite of modern algebraic manipulators provide definite help in handling series expansion and reversion, the procedure described in the previous section can be quite awkward. Then, it emerges the question if the expanded solution in Eqs.~(\ref{pendulum:qseries}), (\ref{pendulum:Qseries}) and (\ref{pendulum:hamSeries}) could be constructed directly. The answer is in the affirmative, and dealing with elliptic functions and series reversion can be totally avoided by setting the rotation regime of the pendulum as a perturbed rotor. Indeed, since previous expansions are based in the convergence of Eq.~(\ref{pendulum:kThetap}), the function $\epsilon\equiv\epsilon(\Theta')$ must be small enough. 
Hence, neglecting the constant summand $mgl$, Eq.~(\ref{pendulum:ham}) can be written as
\begin{equation} \label{pendulum:hame}
\mathcal{H}=\frac{\Theta^2}{2ml^2}\left(1-\frac{2m^2gl^3}{\Theta^2}\cos\theta\right),
\end{equation}
and in those cases in which $2m^2gl^3/\Theta^2\ll1$, that is
\[
mgl\ll\frac{\Theta^2}{2ml^2}
\]
the reduction of the Hamiltonian (\ref{pendulum:hame}) can be achieved as follows.
\par

\subsection{Deprit's triangle}
First, Eq.~(\ref{pendulum:hame}) is written
\[
\mathcal{H}=\frac{\Theta^2}{2ml^2}-\kappa\cos\theta,
\]
where the physical parameter $\kappa=mgl$ is small.\footnote{In fact, one can always choose the units so that $\Theta^2/(2ml^2)$ is of order 1 and $\kappa\ll1$.} Then, the pendulum Hamiltonian can be written as the Taylor series
\begin{equation} \label{pendulum:HTS}
\mathcal{H}=\sum_{n\ge0}\frac{\kappa^n}{n!}H_{n,0}(\theta,\Theta),
\end{equation}
with the coefficients
\begin{eqnarray}
H_{0,0} &=& \frac{\Theta^2}{2ml^2}, \\
H_{1,0} &=& -\cos\theta, \\ \label{Hn0}
H_{n,0} &=& 0 \quad (n>1).
\end{eqnarray}
The reasons for using a double subindex notation will be apparent soon.
\par

The desired transformation $T$ is formally written
\begin{eqnarray*}
\theta &=& \theta(\theta',\Theta';\kappa), \\
\Theta &=& \Theta(\theta',\Theta';\kappa), 
\end{eqnarray*}
which is assumed to be analytic.
Then, when this transformation is applied to Eq.~(\ref{pendulum:hame}), one gets
\[
\Psi\equiv\mathcal{H}(\theta(\theta',\Theta';\kappa),\Theta(\theta',\Theta';\kappa);\kappa),
\]
which is expanded as a Taylor series in the new variables
\begin{equation} \label{psi}
\Psi=\sum_{n\ge0}\frac{\kappa^n}{n!}H_{0,n}, \qquad
H_{0,n}=\left.\frac{\mathrm{d}^n\Psi}{\mathrm{d}\kappa^n}\right|_{\kappa=0},
\end{equation}
where the required derivatives with respect to the small parameter $\kappa$ are computed by the chain rule
\begin{equation} \label{chainrule}
\frac{\mathrm{d}\Psi}{\mathrm{d}\kappa}=
\frac{\partial\mathcal{H}}{\partial\theta}\frac{\mathrm{d}\theta}{\mathrm{d}\kappa}
+\frac{\partial\mathcal{H}}{\partial\Theta}\frac{\mathrm{d}\Theta}{\mathrm{d}\kappa}
+\frac{\partial\mathcal{H}}{\partial\kappa}
\end{equation}

We want the transformation to be canonical and explicit, and hence impose that $T$ is derived from some generating function $\mathcal{W}\equiv\mathcal{W}(\theta,\Theta;\kappa)$. In particular, it is imposed that $T$ be a Lie transform, so that it is defined by the solution of the differential system \cite{LichtenbergLieberman1992,MeyerHallOffin2009}
\begin{equation} \label{lds}
\frac{\mathrm{d}\theta}{\mathrm{d}\kappa}=\frac{\partial\mathcal{W}}{\partial\Theta},
\qquad
\frac{\mathrm{d}\Theta}{\mathrm{d}\kappa}=-\frac{\partial\mathcal{W}}{\partial\theta},
\end{equation}
for the initial conditions $\kappa=0$,
\begin{eqnarray*}
\theta_0 &=& \theta(\theta',\Theta';0)=\theta', \\
\Theta_0 &=& \Theta(\theta',\Theta';0)=\Theta'.
\end{eqnarray*}
The existence of the solution of Eq.~(\ref{lds}) in a neighborhood of the origin, is guaranteed by the theorem of existence and uniqueness 
of the ordinary differential equations. Besides, since the transformation is computed as the solution of a Hamiltonian system, its 
canonicity is also guaranteed.
\par

Then, taking into account Eq.~(\ref{lds}), Eq.~(\ref{chainrule}) is rewritten as
\begin{equation} \label{chainpoiss}
\frac{\mathrm{d}\Psi}{\mathrm{d}\kappa}=
\frac{\partial\mathcal{H}}{\partial\theta}\frac{\partial\mathcal{W}}{\partial\Theta}
-\frac{\partial\mathcal{H}}{\partial\Theta}\frac{\partial\mathcal{W}}{\partial\theta}
+\frac{\partial\mathcal{H}}{\partial\kappa}
=\{\mathcal{H},\mathcal{W}\}+\frac{\partial\mathcal{H}}{\partial\kappa},
\end{equation}
where $\{\mathcal{H},\mathcal{W}\}$ is the Poisson bracket of $\mathcal{H}$ and $\mathcal{W}$.

Replacing $\mathcal{W}$ in Eq. (\ref{chainpoiss}) by its Taylor series expansion
\begin{equation}\label{ltW}
\mathcal{W}=\sum_{n\ge0}\frac{\kappa^n}{n!}\,{W}_{n+1}(\theta,\Theta),
\end{equation}
where the subindex $n+1$ is now used for convenience in subsequents derivations of the procedure, after straightforward manipulations, one arrives to the recurrence
\begin{equation} \label{ltdeprittriangle}
H_{n,q+1}=H_{n+1,q}+
\sum_{0\le{i}\le{n}}{n\choose{i}}\,\{H_{n-i,q};{W}_{i+1}\}.
\end{equation}

Finally, since the terms $H_{0,n}$ needed in Eq.~(\ref{psi}) are evaluated at $\kappa=0$, the original variables are replaced by corresponding prime variables. Full details on the derivation of Eq.~(\ref{ltdeprittriangle}), may be found in Deprit's original reference \cite{Deprit1969} or in modern textbooks in celestial mechanics.
\par

The recurrence in Eq.~(\ref{ltdeprittriangle}) is customarily known as Deprit's triangle. Indeed, the computation of the term $H_{0,n}$ requires the computation of all intermediate terms in a ``triangle'' of ``vertices''  $H_{0,0}$, $H_{0,n}$, and $H_{n,0}$. For instance, for $n=4$ the recurrence in Eq.~(\ref{ltdeprittriangle}) results in a triangle made of:
\begin{center}
\begin{tabular}{lllllllllll}
& & & & ${H}_{0,0}$ \\[1.5ex]
& & & ${H}_{1,0}$ & & ${H}_{0,1}$ \\[1.5ex]
& & ${H}_{2,0}$ & & ${H}_{1,1}$ & & ${H}_{0,2}$ \\[1.5ex]
& ${H}_{3,0}$ & & ${H}_{2,1}$ & & ${H}_{1,2}$ & & ${H}_{0,3}$ \\[1.5ex]
${H}_{4,0}$ & & ${H}_{3,1}$ & & ${H}_{2,2}$ & & ${H}_{1,3}$ & & ${H}_{0,4}$ \\
\end{tabular}
\end{center}
\par

Deprit's triangle is not constrained to the case of one degree of freedom Hamiltonians and generally applies for the transformation of any function $F\equiv{F}(x,X;\kappa)$ where $x\in\mathrm{I\!\!R}^n$, $X\in\mathrm{I\!\!R}^n$ are coordinates and their conjugate momenta, respectively. In particular, when the generation function is given by its Taylor series expansion, it is easily checked that the solution of the differential system (\ref{lds}) is equivalent to applying Deprit's triangle to the functions
\begin{equation} \label{teqs}
\theta=\sum_{n\ge0}\frac{\kappa^n}{n!}\theta_{n,0}(\theta,\Theta), \qquad
\Theta=\sum_{n\ge0}\frac{\kappa^n}{n!}\Theta_{n,0}(\theta,\Theta),
\end{equation}
where $\theta_{0,0}=\theta$, $\Theta_{0,0}=\Theta$, and $\theta_{n,0}=\Theta_{n,0}=0$ for $n>0$.
\par

\subsection{Perturbations by Lie transforms}

Deprit's triangle provides an efficient and systematic way of computing higher orders of a transformation when the generating function is known. However, it is precisely the generating function that is needed to be computed with the only condition that the transformation reduces the pendulum Hamiltonian to a function of only the (new) momentum. Therefore, a new Hamiltonian fulfilling this requirement must be also constructed. The whole procedure of constructing the action-angle Hamiltonian and computing the corresponding generating function can be done in an iterative way as follows.

The construction of $\Phi$ is an iterative procedure based on Deprit's triangle. At each order $n$ Eq.~(\ref{ltdeprittriangle}) is 
rearranged in the form
\begin{equation}\label{lthomological}
\{{W}_n,H_{0,0}\}+H_{0,n}=\widetilde{H}_n,
\end{equation}
which is dubbed as the \textit{homological} equation, where 
\begin{itemize}
\item $\widetilde{H}_n$ is known from previous computations
\item $H_{0,n}$ is chosen in agreement with some simplification criterion such that
\item ${W}_n$ is obtained as a solution of the partial differential equation (\ref{lthomological})
\end{itemize}
Commonly, the simplification criterion for choosing $H_{0,n}$ is eliminating cyclic variables in order to reduce the number of degrees of freedom of the new, transformed Hamiltonian after truncation to $O(\kappa^n)$. This is exactly the case required for the reduction of the pendulum Hamiltonian.
\par

In practice, the procedure starts by reformulating all the functions involved in the homological equations in the prime variables, so that the new Hamiltonian is directly obtained in the correct set of variables.
\par

\subsection{The pendulum as a perturbed rotor}

We start from Eqs.~(\ref{pendulum:HTS})--(\ref{Hn0}). For any term $n$ of the generating function,
\[
\{{W}_n,\mathcal{H}_{0,0}\}=-\frac{\Theta'}{ml^2}\,\frac{\partial{W}_n}{\partial\theta'}.
\] 
\par

The first order of the homological equation is
\begin{equation} \label{homo1}
\frac{\Theta'}{ml^2}\,\frac{\partial{W}_1}{\partial\theta'}=H_{0,1}-\widetilde{H}_1,
\end{equation}
with $\widetilde{H}_1=H_{1,0}$. Then, $H_{0,1}$ is chosen by averaging terms depending on the cyclic variable $\theta'$
\[
H_{0,1}=\frac{1}{2\pi}\int_0^{2\pi}\widetilde{H}_1\,\mathrm{d}\theta'=0,
\]
and ${W}_1$ is solved from Eq.~(\ref{homo1}) by quadrature
\[
{W}_1=\frac{ml^2}{\Theta'}\int(H_{0,1}-\widetilde{H}_1)\,\mathrm{d}\theta'=-\frac{ml^2}{\Theta'}\sin\theta'.
\]
\par

Once ${W}_1$ is obtained, the first order of the transformation is calculated by application of Deprit's triangle to Eq.~(\ref{teqs}) 
with $x\equiv\theta$ and $X\equiv\Theta$. It is obtained
\begin{eqnarray*}
\theta_{0,1} &=& \{\theta',W_1\}=\frac{\epsilon}{\kappa}\sin\theta', \\
\Theta_{0,1} &=& \{\Theta',W_1\}=\Theta'\frac{\epsilon}{\kappa}\cos\theta',
\end{eqnarray*}
where the abbreviation $\epsilon=m^2gl^3/\Theta'^2$ is used, in agreement with Eq.~(\ref{pendulum:epsilonThetap}).

\par

The second order of the homological equation is
\begin{equation} \label{homo2}
\frac{\Theta'}{ml^2}\,\frac{\partial{W}_2}{\partial\theta}=H_{0,2}-\widetilde{H}_2,
\end{equation}
where $\widetilde{H}_2$ is computed from Deprit's triangle
\[
\widetilde{H}_2=\{H_{1,0},{W}_1\}=\frac{ml^2}{\Theta'^2}\sin^2\theta',
\]
$H_{0,2}$ is chosen by averaging
\[
H_{0,2}=\frac{1}{2\pi}\int_0^{2\pi}\widetilde{H}_2\,\mathrm{d}\theta'=\frac{ml^2}{2\Theta'^2}=\frac{\Theta'^2}{2ml^2}\frac{\epsilon^2}{\kappa^2},
\]
and ${W}_2$ is solved from Eq.~(\ref{homo2}) by quadrature
\[
{W}_2=\frac{ml^2}{\Theta'}\int(H_{0,2}-\widetilde{H}_2)\,\mathrm{d}\theta'=-\frac{m^2l^4}{4\Theta'^3}\sin2\theta'.
\]
Then, a new application of Deprit's triangle to the variables provides the second order of the transformation 
\begin{eqnarray*}
\theta _{0,2} &=& \{\theta_{0,1},W_1\}+\{\theta',W_2\}=\frac{1}{4}\frac{\epsilon^2}{\kappa^2}\sin2\theta', \\
\Theta _{0,2} &=& \{\Theta_{0,1},W_1\}+\{\Theta',W_2\}=\Theta'\frac{\epsilon^2}{\kappa^2}\left(\frac{1}{2}\cos2\theta'-1\right),
\end{eqnarray*}
\par

New iterations of the procedure lead to the third order
\begin{eqnarray*}
H_{0,3} &=& 0, \\
\theta _{0,3} &=& \frac{\epsilon^3}{\kappa^3}\left(\frac{33}{8}\sin\theta'+\frac{1}{8}\sin3\theta'\right), \\
\Theta _{0,3} &=& \Theta'\frac{\epsilon^3}{\kappa^3}\left(\frac{9}{8}\cos\theta'+\frac{3}{8}\cos3\theta'\right),
\end{eqnarray*}
the fourth order,
\begin{eqnarray*}
H_{0,4} &=& \frac{\Theta'^2}{2ml^2}\,\frac{15}{4}\frac{\epsilon^4}{\kappa^4}, \\
\theta _{0,4} &=& \frac{\epsilon^4}{\kappa^4} \left(\frac{9}{2}\sin2\theta'+\frac{3}{32}\sin4\theta'\right), \\
\Theta _{0,4} &=& \Theta'\frac{\epsilon^4}{\kappa^4}\left(6\cos2\theta'+\frac{3}{8}\cos4\theta'-\frac{45}{4}\right),
\end{eqnarray*}
the fifth order,
\begin{eqnarray*}
H_{0,5} &=& 0, \\
\theta _{0,5} &=& \frac{\epsilon^5}{\kappa^5}\left(\frac{3705}{32}\sin\theta'+\frac{45}{8}\sin3\theta'+\frac{3}{32}\sin5\theta'\right), \\
\Theta _{0,5} &=& \Theta'\frac{\epsilon^5}{\kappa^5}\left(\frac{585}{32}\cos\theta'+\frac{105}{8}\cos3\theta'+\frac{15}{32}\cos5\theta'\right),
\end{eqnarray*}
and so on.

Now, it is simple to check that the new Hamiltonian $\mathcal{K}=\sum_{n\ge0}(\kappa^n/n!)H_{0,n}$ matches Eq.~(\ref{pendulum:hamSeries}). Besides, rearranging the perturbation series as Fourier series
\begin{eqnarray*}
\theta &=& \sum_{n\ge0}\frac{\kappa^n}{n!}\theta_{0,n}(\theta',\Theta')=\sum_{n\ge0}s_{0,n}(\Theta')\sin(n\theta'), \\
\Theta &=& \sum_{n\ge0}\frac{\kappa^n}{n!}\Theta_{0,n}(\theta',\Theta')=\sum_{n\ge0}c_{0,n}(\Theta')\cos(n\theta'),
\end{eqnarray*}
it is simple to check that they match Eqs.~(\ref{pendulum:qseries}) and (\ref{pendulum:Qseries}), respectively.

\section{Conclusion}

Expanding the transformation to action-angle variables of the simple pendulum solution as a Fourier series, is a cumbersome task because of the elliptic integrals and functions on which the solution relies upon, on the one hand, and the implicit character of the transformation, on the other. Alternatively, these Fourier series expansions can be computed by the Lie transforms method. The latter is straightforward, avoids dealing with special functions, provides the transformation series explicitly without need of series reversion, and is amenable of automatic programming by machine, thus allowing one to compute higher orders of the solution. 
\par

%
%

\end{document}